\documentstyle[twocolumn,aps,epsf]{revtex}

\begin{document}
\draft
\preprint{HEP/123-qed}
\title{Quantum Optimization for Combinatorial Searches}
\author{C. A. Trugenberger}
\address{InfoCodex SA,\\
av. Louis-Casai 18, CH-1209 Gen\`eve, Switzerland}
\address{e-mail: ca.trugenberger@InfoCodex.com}
\date{\today}
\maketitle
\begin{abstract}
I propose a ``quantum annealing" heuristic
for the problem of combinatorial search among
a frustrated set of states characterized by a cost function to be minimized.
The algorithm is probabilistic, with postselection of the measurement result.
A unique parameter playing the role of an effective temperature governs 
the computational load and the overall quality of the optimization. Any level of
accuracy can be reached with a computational load independent of 
the dimension {\it N} of the search set by choosing the effective temperature
correspondingly low. This is much better than classical search heuristics, which
typically involve computation times growing as powers of log({\it N}). 
\end{abstract}

\narrowtext
Quantum computation \cite{reviews}
has recently attracted a widespread interest since it appears
more powerful than its classical counterpart. On one side,
quantum algorithms can be much faster than classical ones. Such algorithms can
tackle problems in new complexity classes which are inaccessible (in polynomial
time) to classical Turing machines, the paramount examples being
Shor's factoring algorithm \cite{shor} and Grover's search algorithm  
\cite{grover}. On the other side, quantum mechanics offers a framework for 
associative memories with exponential 
storage capacity \cite{trugenberger,ezhov}.

The original amplitude amplification via Grover's algorithm \cite{grover} and its 
generalization \cite{generalization} 
are appropriate for decision problems with a simple test
of whether a state satisfies a condition or not, i.e. for searching for ``known
states". On the other hand, a large class of search problems involves such a
quantity of constraints that no single state satisfies them all, i.e. the system
is frustrated \cite{toulouse}. In such combinatorial searches one associates
with each state a cost and the goal is to find the minimum cost state: this is
the problem of optimization.

Most such optimization problems are hard, in the sense that they seem not to be
solvable exactly with a computing effort bounded by a polynomial of the number
$n$ of bits necessary to encode an instance. 
To overcome this hurdle, heuristic methods have been developed which have
computational requirements proportional to small powers of $n$. The price for
this speed-up is that these methods are not guaranteed to find the true 
minimum-cost state but provide only a near-optimal solution. Some of the best
known classes of such optimization heuristics are simulated annealing  
\cite{sa} and genetic algorithms \cite{ga}.

In order to generalize Grover's result to quantum combinatorial searches a
quantum optimization algorithm is needed. Such an algorithm has been proposed by
D\"urr and Hoyer \cite{durr}. It has almost sure success in bounded time ${\rm O}
(\sqrt{N}) $, where $N$ is the dimension of the set to be searched.
Another approach to quantum optimization is, instead, to construct quantum analogues
of classical heuristic algorithms. In this case one might expect both a speed-up of the
above exact quantum result due to the approximate nature of the procedure and an
improvement upon the behaviour of classical counterparts due to the quantum nature 
of the algorithm. In this paper I will show that this is indeed the case.

A first proposal in this direction 
was recently made in \cite{hogg}, where a deterministic quantum optimization
was presented. Quantum algorithms are deterministic if only unitary operations
are performed. On the other hand there are also probabilistic algorithms, in
which intermediate measurements are performed in addition to unitary operations, with
postselection of the measurement result. These are called probabilistic since
the desired result is produced only with a certain probability and repetitions
are thus necessary, a paramount example being probabilistic cloning \cite{pcm}.
Here I will propose a probabilistic quantum optimization algorithm. 

The first step is to 
encode the elements of the search set on states of $n$ qbits (quantum bits).
Let's assume there are $N=O\left(2^n\right)$ such elements, 
corresponding each to a binary number
$I^k$ between 1 and $2^n$. One can then use the algorithm of ref. 
\cite{ventura}, or its simplified version presented in \cite{trugenberger},
to store a superposition of all instances in a single quantum state of 
$n$ qbits:
\begin{equation}
|S\rangle = {1\over \sqrt{N}} \ \sum_{k=1}^N |I^k\rangle \ .
\label{a}
\end{equation}

Next one must identify the cost
function and represent it as a function $C\left( q_1, \dots, q_n\right)$ of the
$n$ bits. The imaginary exponential of this function has then to be realized
with quantum gates \cite{reviews} as a unitary operator on the state $|S
\rangle$. In general this poses problems for the feasibility of quantum
optimization algorithms since it is not guaranteed that ${\rm exp}(iC)$ can
be realized with only simple gates involving few qbits. To avoid this problem
I will assume here that the cost function can be represented by a truncated
expansion
\begin{equation}
C\left( q_1, \dots, q_n\right) = \sum_{k=1}^m \ \sum_{i_1\ne \dots \ne i_k}
C^k_{i_1 \dots i_k}\left( q_{i_1}, \dots, q_{i_k} \right) \ ,
\label{b}
\end{equation}
of terms involving at most $m$ bits. This is actually not a big restriction,
since most combinatorial search problems admit a representation of this
type. As an example I mention the random graph partitioning problem of dividing
an even number $V$ of vertices pairwise connected with probability $p$ by a set $E$
of edges into two sets $V_1$ and $V_2$ of equal size, such that 
the number of edges joining the two sets is minimal. As shown in \cite{anderson},
the cost function given by the number of edges joining $V_1$ and $V_2$ can be
represented as
\begin{equation}
C= {V(V-1)p\over 4} - {1\over 2J} \sum_{i<j} J_{ij} s_i s_j + {\lambda\over 2}
\left( \sum_i s_i \right) ^2 \ ,
\label{adda}
\end{equation}
where $s_i=2q_i-1=1$ if vertex $i \in V_1$ and $s_i = -1$ if vertex $i \in V_2$,
$J_{ij}=J$ if edge $(i,j) \in E$ and $J_{ij} = 0$ otherwise. The last term is a soft
implementation of the antiferromagnetic constraint $\sum_i s_i = 0$.

I will denote by $C_{\rm min}$ and $C_{\rm max}$ strict lower and upper bounds
for the cost function,
\begin{equation}
C_{\rm min} < C\left( q_1, \dots , q_n \right) < C_{\rm max}\ .
\label{newf}
\end{equation}
The unitary operator necessary for quantum optimization is then
\begin{eqnarray}
U &&= {\rm exp} \left( i{\pi \over 2} C_{\rm nor} \left( q_1, \dots, q_n \right)
\right) \ ,
\nonumber \\
C_{\rm nor} \left( q_1, \dots, q_n \right) &&= {{C \left( q_1, \dots, q_n \right)
-C_{\rm min}} \over {C_{\rm max}-C_{\rm min}}} \ .
\label{c}
\end{eqnarray}
For a cost function of the form (\ref{b}), this operator 
can be realized on $|S\rangle$ as
\begin{equation}
U = \prod_{k=1}^m \ \prod_{i_1\ne \dots \ne i_k} \ G^k_{i_1\dots i_k} \ ,
\label{e}
\end{equation}
where $G^k$ are diagonal $k$-qbit gates given by
\begin{eqnarray}
G^k &&= {\rm diag} \left( {\rm e}^{i{\pi \over 2}
C^k_{\rm 
nor}\left( 0_1,\dots ,0_k
\right)}, \dots ,{\rm e}^{i{\pi \over 2}C^k_{\rm nor}\left( 1_1,\dots ,1_k
\right)} \right) \ ,
\nonumber \\
&&C^k_{\rm nor} \left( q_1, \dots , q_k\right) = 
{{C^k \left( q_1, \dots, q_k \right)
-{1\over K} C_{\rm min}} \over {C_{\rm max}-C_{\rm min}}} \ ,
\nonumber \\
&&K  = \sum_{k=1}^m \ k! \left({n\atop k}\right) \ ,
\label{f}
\end{eqnarray}
and the subscripts denote generically the qbits on which operators are applied.
For $m$ not too large, the unitary operator $U$ is thus feasible, since it can
be realized with simple quantum gates involving only few qbits. 

Having introduced the two building blocks of quantum optimization, the search
state $|S\rangle$ and the unitary operator $U$, I will proceed to the
description of the optimization algorithm proper. This involves, in addition to
the $n$ qbits in state $|S\rangle$, also a second register with $b$ control
qbits $|c_1, \dots , c_b\rangle$, all initially in state $|0\rangle$. The
initial quantum state is thus given by
\begin{equation}
|\psi_0\rangle = {1\over \sqrt{N}} \ \sum_{k=1}^N 
\ |I^k; 0_1,\dots , 0_b\rangle \ .
\label{g}
\end{equation}
Applying the Hadamard gate \cite{reviews}
\begin{equation}
H = {1\over \sqrt{2}} \ \left( \matrix{1 & 1\cr
1 & -1\cr} \right) 
\label{h}
\end{equation}
to the first control qbit one obtains
\begin{eqnarray}
|\psi_1\rangle &&= {1\over \sqrt{2N}} \ \sum_{k=1}^N \ |I^k; 0_1,\dots , 0_b\rangle 
\nonumber \\
&&+ {1\over \sqrt{2N}} \ \sum_{k=1}^N \ |I^k; 1_1, \dots , 0_b \rangle \ .
\label{i}
\end{eqnarray}

At this point I need to introduce the controlled gate
\begin{equation}
U^{\pm}_{cS} = |0_c\rangle \langle 0_c| \otimes U_S + |1_c\rangle \langle 1_c|
\otimes {U_S}^{-1} \ ,
\label{l}
\end{equation}
which realizes on the search state $|S\rangle$ the unitary transformation
$U$ if the control qbit $c$ is in state $|0\rangle$ and the unitary
transformation $U^{-1}$ if control qbit $c$ is in state $|1\rangle$. This can
be realized as
\begin{eqnarray}
&&U^{\pm}_{cS} = \prod_{k=1}^m \ \prod_{i_1\ne \dots \ne i_k}
\left( C {G^k}^{-2} \right) _{ci_1 \dots i_k} \ G^k_{i_1\dots i_k}\ ,
\nonumber \\
&&\left( C {G^k}^{-2} \right) _{ci_1 \dots i_k} =
|0_c\rangle \langle 0_c| \otimes 1_{i_1\dots i_k} + |1_c\rangle \langle 1_c|
\otimes {G^k}^{-2}_{i_1\dots i_k} \ ,
\label{m}
\end{eqnarray}
where $C {G^k}^{-2}$ is the standard controlled ${G^k}^{-2}$ gate, 
realized only if the control qbit is in 
state $|1\rangle$ \cite{reviews}. This shows that also $U^{\pm}$ is feasible,
since at most $(m+1)$-qbit gates are involved.

Applying $U^{\pm}_{c_1S}$ to $|\psi _1\rangle$ gives
\begin{eqnarray}
|\psi _2\rangle &&= {1\over \sqrt{2N}} \ \sum_{k=1}^N {\rm e}^{i{\pi \over 2}
C_{\rm nor}\left( I^k \right)} \ |I^k;0_1,\dots ,0_b\rangle 
\nonumber \\
&&+{1\over \sqrt{2N}} \ \sum_{k=1}^N {\rm e}^{-i{\pi \over 2}
C_{\rm nor}\left( I^k \right)} \ |I^k;1_1,\dots ,0_b\rangle \ .
\label{n}
\end{eqnarray}
Note that the cost function determines the relative phase of the search states. This
is a generalization of the bit flip in Grover's algorithm \cite{grover},
where the cost function takes only the values $0$ or $1$ (or a constant). The following
concentration of amplitude on low-cost states, instead, is a totally different mechanism
than the diffusion operator acting in Grover's algorithm.

Applying again the Hadamard gate $H$ to the first control qbit $c_1$ one
obtains finally
\begin{eqnarray}
|\psi _3\rangle &&= {1\over \sqrt{N}} \ \sum_{k=1}^N {\rm cos}
\left( {{\pi \over 2}
C_{\rm nor}\left( I^k \right)}\right) \ |I^k;0_1,\dots ,0_b\rangle 
\nonumber \\
&&+{1\over \sqrt{N}} \ \sum_{k=1}^N {\rm sin} \left( {{\pi \over 2}
C_{\rm nor}\left( I^k \right)}\right) \ |I^k;1_1,\dots ,0_b\rangle \ .
\label{p}
\end{eqnarray}
Essentially, the steps $H_cU^{\pm}_{cS}H_c$ represent the quantum equivalent of
the classical evaluation of the cost function on a search state. Here, the cost
function is evaluated on all possible search states at the same time.

This evaluation has now to be repeated $b$ times 
for all control qbits $c_1$ to $c_b$, giving the final result
\begin{eqnarray}
|\psi_{\rm fin}\rangle = {1\over \sqrt{N}} \sum_{k=1}^N \sum_{i=0}^b
\ &&{\rm cos}^{b-i} \left( {\pi\over 2} C_{\rm nor}\left( I^k\right) \right) \times 
\nonumber \\
&&{\rm sin}^i \left( {\pi\over 2} C_{\rm nor}\left( I^k\right) \right) 
\ \sum_{\left\{ J^i \right\}} |I^k; J^i\rangle ,
\label{q}
\end{eqnarray}
where $\left\{ J^i \right\}$ denotes the set of all binary numbers of
$b$ bits with exactly $i$ bits 1 and $(b-1)$ bits 0. 
Note that the overall effect of the $b$ operators $H_cU^{\pm}_{cS}H_c$
is an {\it amplitude concentration on low-cost search states} for such 
complete states which have a large number of 0 control qbits and an amplitude
concentration on high-cost search states if there are many control qbits with
value 1. This is the core of the deterministic part of the quantum optimization
procedure, which is concluded here.

At this point one proceeds to a measurement of the control register. Given that
the amplitude is most concentrated on low-cost states when all control qbits are
in state $|0\rangle$ one retains the resulting projected state only if the
control register is measured in state $|0_1, \dots ,0_b \rangle $. In general
this entails some repetitions of the deterministic transformation described
above and of the meausrement until this state is obtained. The expected number
of these repetitions is $1/P^0_b$, where $P^0_b$ is the probability that
$|c_1, \dots, c_b\rangle = |0_1, \dots, 0_b\rangle$:
\begin{equation}
P^0_b = {1\over N} \ \sum_{k=1}^N \ {\rm cos}^{2b} 
\left( {\pi\over 2} C_{\rm nor}
\left( I^k\right) \right) \ .
\label{r}
\end{equation}
Once established that the control register is in the desired state one can
proceed to a measurement of the search state $|S\rangle$. This measurement
will yield state $|I^k\rangle$ with a probability
\begin{eqnarray}
P_b\left( I^k \right) &&= {1\over Z} \ {\rm cos}^{2b} 
\left( {\pi\over 2} C_{\rm nor}
\left( I^k\right) \right) \ ,
\label{sa} \\
Z &&= N P^0_b \ .
\label{sb}
\end{eqnarray}
which is peaked on the low-cost states, exactly as desired for an optimization
heuristic. Overall, the quantum optimization procedure consists thus of a 
series of independent trials, each of which returns a set of values for the
control qbits. As soon as one obtains the desired values for these control qbits
one can proceed to measure the search state, with a high probability of
selecting a low-cost state. 

Completing the specification of the algorithm requires selecting a value for the
number $b$ of control qbits. Unfortunately, there is no general rule to select
the best value of $b$, since this depends on the problem at hand. 
It is important, however, to point out that there is a 
generic trade-off between the advantages
and disadvantages of high and low values of $b$. 
Selecting a high value of $b$ clearly enhances the probability of finding the
true minimum-cost state, since the probability distribution $P_b$ becomes
more and more peaked towards the low-cost states:
\begin{equation}
\lim_{b\to \infty} P_b\left( I^k \right) = \delta_{kk_{\rm min}}\ ,
\label{t}
\end{equation}
where $k_{\rm min}$ is the index of the minimum-cost state (assumed for simplicity
to be unique). On the other hand,
however, high values of $b$ make the probability $P^0_b$ of measuring all
control qbits in state $|0\rangle$ lower, increasing thus the expected number
of repetitions of the deterministic part of the algorithm before one obtains
the desired values of the control qbits permitting to proceed. 

For the purpose of comparing with classical algorithms, 
a measure of the computational load of a combinatorial search heuristic can be
taken as the number of times the cost function has to be evaluated.
For classical algorithms this scales usually as $n$ or a small power 
of $n$, in order to obtain good-quality results. Quantum parallelism, instead,
allows one to evaluate the cost function on all instances at the same time. The price
to pay is the expected number $1/P^0_b$ of repetitions of this overall
evaluation in order to obtain the correct state for a measurement. These
repetitions are however harmless, since their number is independent of $n$
for large $n$. This follows from the fact that the probability $P^0_b$ 
in (\ref{r}) has a finite, non-vanishing large-$n$ limit, given that
$0 < C_{\rm nor}\left( I^k \right) < 1$ for all $I^k$. 

As a consequence, the computational load of quantum optimization is determined 
entirely by the parameter $b$.
The role of this parameter can be better understood by examining closer
eqs. (\ref{sa}) and (\ref{sb}). The quantum distribution described by these
equations is equivalent to a canonical Boltzmann distribution with
(dimensionless) {\it effective temperature} 
$t=1/b$ and (dimensionless) energy levels given by
\begin{equation}
E^k = -2 \ {\rm log} \ {\rm cos} \left( {\pi \over 2} C_{\rm nor} \left( I^k
\right) \right) \ ,
\label{z}
\end{equation}
with $Z$ in eq. (\ref{sb}) playing the role of the partition function.
The relation between the unbounded, positive energies $E$ entering the effective
thermal distribution and the bounded, normalized cost function simplifies in 
the two limits of low and high cost:
\begin{eqnarray}
E \ &&\simeq {\pi ^2 \over 4} \ C^2_{\rm nor} 
 \quad \quad \quad \quad \quad \quad \quad C_{\rm nor} \ll 1 \ ,
\nonumber \\
E \ &&\simeq \ {\rm log} \ {4\over \pi^2 \left( 1-C_{\rm nor} \right) ^2 } 
\quad \quad
1-C_{\rm nor} \ll 1 \ .
\label{za}
\end{eqnarray}
As expected, the deviation is largest for high-cost configurations: the
logarithmic transformation maps the region near the bound 1 of $C_{\rm nor}$
to the positive axis for $E$. 

The appearance of an effective thermal distribution highlights the analogy
between this quantum optimization procedure and simulated annealing. In this
classical search heuristic one approaches a low-cost state by lowering the
temperature in a simulated thermal ensemble generated by an appropriate
number of Monte-Carlo steps. In the quantum optimization algorithm
proposed here one
obtains a low-cost configuration by choosing the effective temperature
$t=1/b$ low enough. This can be tuned by adding an appropriate number of control
qbits. 

As in classical simulated annealing, one can use 
the effective thermal ensemble to derive the
average behaviour of the optimization. As first pointed out in \cite{sa},
this is a better measure of the heuristic performance than a worst-case analysis
when the size of the problem becomes large. Following the classical approach
\cite{parisi} I shall thus concentrate on the free energy F defined by
\begin{equation}
Z = 2^n \ e^{-bF(b)} = Z(b=0) \ e^{-bF(b)} \ ,
\label{newa}
\end{equation}
where I have chosen a normalization such $e^{-bF(b)}$
describes the deviation of the partition function from its value for $b=0$.
Since $Z/2^n$ possesses a finite, non-vanishing
large-$n$ limit, as explained above, this normalization ensures that
$F(b)$ is intensive, exactly like the energy levels (\ref{za}), and scales as
a constant for large $n$.

The free energy describes the equilibrium of the system at effective temperature
$t=1/b$ and has the usual expression in terms of the internal energy $U$ and the
entropy $S$,
\begin{eqnarray}
F(t) &&= U(t) - tS(t) \ ,
\nonumber \\
U(t) && = \langle E \rangle _t \ ,
\nonumber \\
S(t) &&= {-\partial F(t) \over \partial t} \ .
\label{newg}
\end{eqnarray} 
By inverting equation (\ref{z}) one can then also define an effective cost
function ${\cal C}(t)$ at temperature t:
\begin{eqnarray}
{\cal C}_{\rm nor} (t) &&= {2\over \pi} \ {\rm arccos} \ e^{-{F(t)\over 2}} \ ,
\nonumber \\
{\cal C} (t) &&= C_{\rm min} + \left( C_{\rm max} - C_{\rm min} \right)
\ {\cal C}_{\rm nor} (t) \ .
\label{newb}
\end{eqnarray}
This corresponds exactly to representing the probability (\ref{r}) as
\begin{equation}
P^0_b = {\rm cos}^{2b} \left( {\pi \over 2} \ {\cal C}_{\rm nor} (b) \right) \ ,
\label{newc}
\end{equation}
which can be taken as the primary definition of the effective cost function
${\cal C}$.

At $t=\infty $ all configurations $I^k$ contribute equally to the partition
function and thus ${\cal C}(t=\infty)$ is a non-linear function of
the average cost computed from
equation (\ref{b}). At $t=0$ only the optimal configuration survives in the
partition function and ${\cal C}(t=0)$ is thus exactly the minimum cost,
${\cal C}(t=0) = C_{\rm opt}$. In general one can write
\begin{equation}
{\cal C}(t) = {\cal C}(\infty ) - \Delta (t)\ ,
\label{newd}
\end{equation}
with $\Delta (t)$ representing the average gain due to the optimization at
effective temperature $t$. A measure of the accuracy of the optimization is then
given by 
\begin{equation}
{\Delta (t) \over {{\cal C}(\infty ) - {\cal C}(0)}} \ .
\label{newe}
\end{equation}
By construction this is independent of $n$ to leading order. Any level of
accuracy can thus be reached by tuning the effective temperature $t$ to values
low enough, independently of $n$. The computational load is determined by
$b=1/t$ and is thus also independent of $n$. In other words one can define a
proper thermodynamics by taking the limit $n\to \infty$ in equations (\ref{sa})
and (\ref{sb}). In this limit one is left with the parameter $t$ alone: this
determines both the accuracy via (\ref{newe}) and the computational load via
$b=1/t$. 

The quantum optimization algorithm described here represents a huge improvement
upon the exact quantum search for the minimum \cite{durr}, which has a
computational load of $\sqrt{N}$, with $N=2^n$. The price to pay is the
approximate character of the solution. It is this approximation which allows the
improvement without violating known bounds for exact 
quantum results \cite{bounds}. My quantum optimization procedure is also much
faster than classical search heuristics, which typically involve computation
times growing like small powers of log({\it N}). The reason here lies in quantum
parallelism. This is seen most explicitly by comparing with classical simulated
annealing: in this algorithm a thermal distribution has to be simulated by
computation at each temperature. This involves long relaxation times which
constitute already one source of the log({\it N}) computational load. Due to
quantum parallelism, instead, the whole thermal distribution is generated in one
go.

Let me conclude by mentioning a particularly interesting case. This arises when
the effective thermal distribution undergoes a (first order) 
phase transition at $t=t_{\rm
cr}$. In this case one has ${\cal C}(t) = {\cal C}(\infty )$ for 
$t > t_{\rm cr}$, while for $t< t_{\rm cr}$ the system is frozen into a 
``quantum solid state" with ${\cal C}(t)$ rapidly approaching its limiting value
${\cal C}(0)$. An example of this phenomenon in the framework of quantum pattern
recognition will appear in a forthcoming publication. 


\end{document}